\documentclass[11pt]{revtex4}

\usepackage{hyperref}
\usepackage{amsmath,amsfonts,amssymb}
\hypersetup{dvips,dvipdfm,colorlinks=true,urlcolor=magenta,filecolor=magenta,linktoc=page,citecolor=red,linkcolor=blue,bookmarks=true}
\usepackage{graphicx,epstopdf}

\begin{document}

\title{Quantum cosmology for non-minimally coupled scalar field in FLRW space--time: A symmetry analysis}

\author{Sourav Dutta$^1$\footnote {sduttaju@gmail.com}}
\author{Muthusamy Lakshmanan $^2$\footnote {lakshman.cnld@gmail.com}}
\author{Subenoy Chakraborty$^3$\footnote {schakraborty.math@gmail.com}}
\affiliation{$^1$Department of Pure Mathematics, University of Calcutta, Ballygunge Science College, 35, Ballygunge Circular Rd, Ballygunge, Kolkata, West Bengal 700019\\
	$^2$ Centre for Nonlinear Dynamics, Bharathidasan University, Tiruchirapalli - 620 024, India\\
	$^3$Department of Mathematics, Jadavpur University, Kolkata-700032, West Bengal, India\\}


\begin{abstract}
The present work deals with quantum cosmology for non-minimally coupled scalar field in the background of FLRW space--time model. The Wheeler-DeWitt equation is constructed and symmetry analysis is carried out. The Lie point symmetries are related to the conformal algebra of the minisuperspace while solution of the Wheeler-DeWitt equation is obtained using conserved currents of the Noether symmetries.
	
\end{abstract}

\maketitle

\section{Introduction}
The recent observational evidences \cite{r1}--\cite{r7} that Universe has been going through an accelerated phase of expansion is contradictory to standard cosmology. However, the observational evidences have been nicely accommodated in the framework of general relativity by incorporating an exotic matter \cite{r8}--\cite{r10} having large negative pressure, namely the dark energy (DE). The simplest and the automatic choice for the DE candidate is the cosmological constant \cite{r11}--\cite{r16}. But this choice of DE model \cite{n1}--\cite{n5} is not acceptable to the cosmologists due to its two severe problems: extreme fine tuning problem and the coincidence problem \cite{r17}. So cosmologists have been trying for dynamical dark energy models having a variable equation of state (with negative energy) \cite{r18} to accommodate the observational evidences. In the present work, the dynamical dark energy is chosen as a non-minimally coupled scalar field having self interacting potential \cite{r19}--\cite{r21.3}. The evolution equations of the present cosmological model are highly non-linear and coupled second order differential equations. It is hard to find exact analytic solution using the usual techniques. Here symmetry analysis technique will be imposed to find analytic solutions, particularly in the quantum domain.\\

There are significant developments in the symmetry analysis over the last century. In particular, symmetry study is now not only confined to the study of global continuous symmetries (namely translation, rotation etc.), but it even deals with local continuous symmetries, particularly local gauge symmetries, internal symmetries to the space--time in cosmology, and permutation symmetry in quantum field theory \cite{r22}, \cite{r23}. Today it is generally believed that symmetries are the key tools in the formulation of fundamental physics: Principle of relativity in Einstein's construction of special relativity, general covariance in general theory of relativity, group theory in quantum field theory and symmetry principles in standard model for particle physics.

In the present context, basic geometrical symmetries (of the space-time) namely Lie point and Noether are very useful in physical problems. Mathematically, Lie point/ Noether symmetries play an important role either to simplify the physical system or to determine the integrability of the system \cite{r24}--\cite{r28.2}. Noether symmetry of a physical system is of much interest because it provides conservation laws for the physical theory. In particular, symmetries of dynamical systems are related to their first integrals for the physical quantities which do not evolve with the system and are related to the fundamental physical quantities, namely energy, linear momentum, angular momentum, etc. \cite{r29}, \cite{r30}. Further, in the context of quantum cosmology, Noether symmetries help one to obtain a typical subset of the general solution of the Wheeler-DeWitt equation having oscillating  behaviors \cite{r31}, \cite{r32}, \cite{r21.1}. Moreover, in the context of minisuperspace criterion they also select equations of classical trajectories \cite{r32}, \cite{r34}. So Noether symmetries act as a bridge to relate classically observable Universe to quantum cosmology.\\

 In the present work, Wheeler-DeWitt (WD) equation in quantum cosmology is constructed for non-minimally coupled scalar field cosmology in the background of FLRW space-time model. Basic geometric symmetries of the space-time namely Lie and Noether symmetries are used to the WD equation as a tool for solving it. Also conformal symmetry has been studied in the context of quantum cosmology. The paper has been organized as follows: The minisuperspace approach in quantum cosmology has been introduced in section II. Section III gives a general description of conformal symmetry. The WD equation is formulated for the present cosmological model in section IV and wave function of the Universe has been evaluated using Lie symmetry of the hyperbolic partial differential equation. A general construction of the Noether symmetry has been presented in section V. Using this Noether symmetry the WD equation is simplified to a great extent and possible solutions have been derived in section VI. Finally, the paper ends with a brief discussion and concluding remarks in section VII. 

\section{Quantum cosmology: The Minisuperspace approach} 
The symmetries in superspace characterize the metric and matter fields while minisuperspaces are restrictions of geometrodynamics of the superspace. Usually in cosmology, physically relevant and interesting models are defined on minisuperspaces. The common and simplest minisuperspace model consists of homogeneous and isotropic metrics and matter fields so that the lapse function is homogeneous i.e., $N=N(T)$ with vanishing shift function. Thus the metric on the four dimensional manifold can be written as 
\begin{equation}
ds^2=-N^2(t) dt^2+h_{ab} (x, t)~ dx^a ~dx^b, \label{q1}
\end{equation} 
so that the Einstein-Hilbert action takes the form
\begin{equation}
A(h_{ab}, N)=\frac{m_p^2}{16 \Pi}\int dt~ d^3 x N\sqrt{h}\Bigg[K_{ab} K^{ab}-k^2+{}^{(3)}R - 2\Lambda\Bigg] , \label{q2}
\end{equation}
where $K_{ab}$ is the extrinsic curvature, $k=K_{ab} h^{ab}$ is the trace of the extrinsic curvature, ${}^{(3)}R$ is the three-space curvature scalar and $\Lambda$ is the cosmological constant.

As the homogeneous three metric $h_{ab}$ is characterized by a finite number of functions $q^{\alpha}(t), \alpha=0, 1, 2, ....(n-1),$ so the above action takes the form:
\begin{equation}
A\bigg(q^{\alpha}(t), N(t)\bigg)=\int^1_0 dt N\Bigg[\frac{1}{2N^2} f_{\alpha \beta} (q) \dot{q}^{\alpha} \dot{q}^{\beta}-V(q)\Bigg] , \label{q3}
\end{equation}
 where $f_{\alpha \beta}$ is the metric on the minisuperspace. This action has the familiar form of a relativistic point particle having self-interaction potential $V(q)$ moving in a $n$-dimensional curved space-time. Now, variation of the action with respect to the field variables $q^{\alpha}(t)$ gives the equation of motion of the (equivalent) relativistic particle as 
 \begin{equation}
 \frac{1}{N} \frac{d}{dt} \Bigg(\frac{\dot{q}^{\alpha}}{N}\Bigg)+\frac{1}{N^2} \Gamma^{\alpha}_{\beta \gamma}~ \dot{q}^{\beta} ~\dot{q}^{\gamma}+f^{\alpha \beta} \frac{\partial V}{\partial q^{\beta}}=0, \label{q4}
 \end{equation}
 where $\Gamma^{\alpha}_{\beta \gamma}$ is the Christoffel symbol, while variation with respect to the lapse function gives the constraint equation 
 \begin{equation}
 \frac{1}{2N^2} f_{\alpha \beta}~\dot{q}^{\alpha}~\dot{q}^{\beta}+V(q) =0. \label{q5}
 \end{equation}
 As a result, the general solution of the above evolution equations contains $(2n-1)$ arbitrary parameters.
 
 We shall now discuss the Hamiltonian dynamics of the system. The canonical Hamiltonian is defined as
 \begin{equation}
 \mathcal{H}_c=p_{\alpha} \dot{q}^{\alpha}-\mathcal{L}=N \Bigg[\frac{1}{2} f^{\alpha \beta} p_{\alpha} p_{\beta}+V(q)\Bigg]\equiv N \mathcal{H}, \label{q6}
 \end{equation}
 where $p_{\alpha}=\frac{\partial \mathcal{L}}{\partial \dot{q}^{\alpha}}=f_{\alpha \beta} \frac{\dot{q}^{\beta}}{N}$ is the momenta canonical to $q^{\alpha}$ and $f^{\alpha \beta}$ is the inverse metric. Using the above expression for $p_{\alpha}$ to the constraint equation (\ref{q5}) gives the Hamiltonian constraint equation:
 \begin{equation}
 \mathcal{H}(q^{\alpha}, p_{\alpha})=\frac{1}{2} f^{\alpha \beta} p_{\alpha} p_{\beta}+V(q)=0.\label{q7}
 \end{equation}
 To proceed with canonical quantization, one has to construct the Wheeler-DeWitt (WD) equation which is nothing but the quantum operator version of the constraint equation (\ref{q7}) operating on a time independent function (known as wave function of the Universe). In the present context the WD  equation is
 \begin{equation}
\hat{\mathcal{H}}(q^{\alpha},-i \frac{\partial}{\partial q^{\alpha}})\psi(q^{\alpha}) =0.\label{q8}
 \end{equation}
 Due to the dependence of $f^{\alpha \beta}$ on $q$ there is an ambiguity related to factor ordering in the above WD equation. However, it may be resolved by demanding that the quantization in minisuperspace is covariant in nature i.e., invariant under the change in fields:$q^{\alpha} \rightarrow \tilde{q}^{\alpha} (q^{\alpha}).$ As a result the Hamiltonian operator becomes
 \begin{equation}
 \hat{\mathcal{H}}\equiv -\frac{1}{2} \bigtriangledown^2+\xi \mathcal{R}+V(q),\label{q9}
 \end{equation}
where the Laplacian operator $\bigtriangledown^2$ and curvature scalar $\mathcal{R}$ are defined over the minisuperspace metric $f_{\alpha \beta}$, and $\xi$ is an arbitrary constant.\\

We shall now discuss the issue of probability measure in quantum cosmology. Usually, for hyperbolic type of partial differential equations (PDE) $\exists$ a conserved current 
  \begin{equation}
\overrightarrow{J}=\frac{i}{2}\big(\psi^{*} \bigtriangledown \psi-\psi \bigtriangledown \psi^{*}\big), \label{q10}
 \end{equation}
 with $\overrightarrow{\bigtriangledown}.\overrightarrow{J}=0$. Here $\psi$ is a solution of the hyperbolic PDE (note that the WD equation is a hyperbolic PDE on minisuperspace). One can define the probability from this conserved current but it is not free from negative probabilities. As a result, a correct probability measure on the minisuperspace can be chosen as 
 \begin{equation}
 dp=|\psi(q^{\alpha})|^2 dV,\label{q11}
 \end{equation}
 where $dV$ is a volume element on minisuperspace.

\section{Conformal symmetry}

There are rich geometrical structures corresponding to conformal invariance. A vector field $X^a$ is a conformal killing vector (CKV) of the metric $g_{ij}$ if
\begin{equation}
 \mathcal{L}_{\overrightarrow{X}} g_{ij}=\chi (x^k) g_{ij}, \label{c1}
\end{equation}
where $\mathcal{L}_{\overrightarrow{X}}$ stands for the Lie derivative of the Lagrangian with respect to the vector field $\overrightarrow{X}$ and $\chi$ is a function of the space. As a particular case if $\chi$ is a non-zero constant (say $\chi_{_0}\neq 0$) i.e., $\mathcal{L}_{\overrightarrow{X}} g_{ij}=\chi_{_0} g_{ij}$ then $\overrightarrow{X}$ is called a homothetic vector field, while for $\chi_{_0} =0$ i.e., $\mathcal{L}_{\overrightarrow{X}} g_{ij}=0$, the vector field $\overrightarrow{X}$ is called a killing vector field.\\\\

Two metrics $g$ and $\bar{g}$ of the same space are said to be conformally related if $\exists$ a function $\xi (x^k)$ such that
\begin{equation}
 \bar{g}_{ij}=\xi^2 (x^k) g_{ij}.\label{c2}
\end{equation}

Now the class of conformal killing vectors form an algebra which is termed as conformal algebra of the metric \cite{r35}. Similarly, the set of homothetic vector fields and the set of killing vector fields also form algebras known as homothetic algebra (HA) and killing algebra (KA) respectively. These two algebras are closed subalgebras of the conformal algebra (CA)
\begin{equation}
 KA \subseteq HA \subseteq CA.\label{c3}
\end{equation}
In an $n$ dimensional manifold ($n> 2$) of constant curvature the dimension of these three algebras are $\frac{(n+1)(n+2)}{2},~\frac{n(n+1)}{2}+1$ and $\frac{n(n+1)}{2}$ respectively. Although, two conformally related metrics have the same conformal algebra but the subalgebras are not the same. However, if $\overrightarrow{X}$ is a conformal killing vector for the conformally related metrics $g$ and $\bar{g}$ having conformal factors $\chi(x^k)$ and $\bar{\chi}(x^k)$ respectively then one has
\begin{equation}
 \bar{\chi}(x^k)=\chi(x^k)+\mathcal{L}_{\overrightarrow{X}}(\ln \xi).\label{c4}
\end{equation}
 As physical systems are mostly described by appropriate Lagrangians it will be interesting to study the notion of conformal Lagrangian. It has been shown by Tsamparlis et al.\cite{r35} that the equation of motion (i.e., the Euler-Lagrange equations) corresponding to two conformal Lagrangians transform covariantly under the conformal transformation provided the total energy (i.e., the Hamiltonian) is zero. Equivalently, systems with vanishing energy are conformally related and corresponding equations of motion are conformally invariant. \\

Subsequently, the authors in reference \cite{r35} have extended the idea of conformally equivalent Lagrangian to scalar field cosmology in general Riemannian space. According to them, a non-minimally coupled scalar field cosmology is equivalent to a minimally coupled scalar field cosmology in a conformally invariant metric, provided: (i) the Lagrangians are conformally related  and (ii) the coupling function is given by $F(\psi)=-(2 \xi^2)^{-1} <0$. However, in the context of quantum cosmology, due to the Hamiltonian constraint the total energy of the system has to be zero and hence one has conformally invariant systems with respect to equations of motion.\\\\
Finally, from the point of view of Noether symmetries the above conformally related physical systems are not identical because the Noether symmetries follow the homothetic algebra of the metric which are distinct for two conformally related metrics.

 \section{Formulation of Wheeler-DeWitt equation in the present cosmological model and the Lie point symmetry}
 
 In the background of flat FLRW space--time
 \begin{equation}
 ds^2=-dt^2+a^2(t) \Bigg[dr^2+r^2 d\Omega_2^2\Bigg], \label{3.1}
 \end{equation}
 where $a(t)$ is a scalar function and $d\Omega_2^2=d\theta^2+\sin^2 \theta~d\phi^2$ is a metric on the unit-2 sphere.
 The Lagrangian for non-minimally coupled scalar field with non-interacting hot dark matter is given by \cite{r21.1}, \cite{r21.2}
 \begin{equation}
 L(a, \dot{a}, \phi, \dot{\phi})=-3a \dot{a}^2+\frac{1}{2} a^3 \lambda(\phi) \dot{\phi}^2-a^3 V(\phi)-\rho_0 a^{3(1-\gamma)}. \label{3.2}
\end{equation}
 The Friedmann equations for the model are given by
 \begin{equation}
 3\frac{\dot{a}^2}{a^2}=\rho_m+\rho_{\phi},\label{3.3}
 \end{equation}
 and 
 \begin{equation}
 2 \frac{\ddot{a}}{a}=-\frac{1}{3}\Bigg[(\rho_m+3p_m)+(\rho_{\phi}+3p_{\phi})\Bigg].\label{3.4}
 \end{equation}
 Here $\lambda(\phi)$ and $V(\phi)$ are respectively the coupling function and potential of the non-minimally coupled scalar field having energy density $\rho_{\phi}$ and thermodynamic pressure $p_{\phi}$ as
 \begin{equation}
 \rho_{\phi}=\frac{1}{2} \lambda(\phi) \dot{\phi}^2+V(\phi),~p_{\phi}=\frac{1}{2} \lambda(\phi) \dot{\phi}^2-V(\phi).\label{3.5}
 \end{equation}
 The hot dark matter has $(\rho_m, p_m)$ as the energy density and thermodynamic pressure having equation of state parameter:~$p_m=(\gamma-1)\rho_m, \gamma$ a constant. The evolution of these two non-interacting matter components are
 \begin{eqnarray}
 \dot{\rho_{\phi}}+ 3H(\rho_{\phi}+p_{\phi})=0,\nonumber
 \end{eqnarray}
 that is
 \begin{eqnarray}
 \lambda(\phi)\ddot{\phi}+\frac{1}{2} \lambda'(\phi) \dot{\phi}^2+3H \lambda(\phi) \dot{\phi}+\frac{dV}{d\phi}=0,\label{3.6}
 \end{eqnarray}
 and
 \begin{equation}
\dot{\rho_m}+3H(\rho_m+p_m)=0,\label{3.7}
 \end{equation}
 which on using the above condition ($p_m=(\gamma-1)\rho_m$) and integration gives
 \begin{equation}
\rho_m=\rho_0 a^{-3\gamma},\label{3.8}
 \end{equation}
 where $H=\frac{\dot{a}}{a}$ is the usual Hubble parameter.\\
 
 Note that equation (\ref{3.6}) can be obtained from the Lagrangian (\ref{3.2}) by variation with respect to the scalar field $\phi$. Also due to the absence of explicit time dependence of the Lagrangian, the Einstein equation (\ref{3.3}) is termed as the Hamiltonian constraint equation.\\
 
 The momenta conjugate to the configuration variables $a$ and $\phi$ are given by
 \begin{equation}
 p_a=\frac{\partial L}{\partial \dot{a}}=-6a \dot{a},~p_{\phi}= \frac{\partial L}{\partial \dot{\phi}}=a^3 \lambda(\phi) \dot{\phi}. \label{3.9}
 \end{equation}
 So the above Hamiltonian constraint in terms of the momenta takes the form:
 \begin{equation}
 \mathcal{H}\equiv -\frac{1}{12} \frac{p_a^2}{a}+\frac{1}{2a^3} \frac{p_{\phi}^2}{\lambda(\phi)}+a^3 V(\phi)+\rho_0 a^{-3(\gamma-1)}=0.\label{3.10}
 \end{equation}
 For the above Hamiltonian, the equivalent Hamilton's equation of motion are
 \begin{eqnarray}
 \dot{a}&=&-\frac{1}{6a} p_a,~\dot{\phi}=\frac{1}{a^3 \lambda(\phi) }p_{\phi}, \nonumber\\
 \dot{p_a}&=&-\frac{1}{12} \frac{p_a^2}{a^2}+\frac{3p_{\phi}^2}{2a^4 \lambda(\phi)}-3a^2 V(\phi)+3(\gamma-1) \rho_0 a^{-3\gamma+2}, \nonumber\\
 \dot{p_{\phi}}&=&-a^3 V'(\phi)+\frac{p_{\phi}^2 \lambda'(\phi) }{2a^3 \lambda^2(\phi)}.\label{3.11}
 \end{eqnarray}
 
The Lagrangian (\ref{3.2}) of the system can be divided into two parts: the kinetic part which consists of the first two terms while the last two terms together are known to constitute the dynamic part. In fact, the kinetic part may be considered as a 2D Riemannian space having the line element
\begin{equation}
ds^2=-6a~ da^2+a^3 \lambda(\phi)~ d\phi^2.\label{3.12}
\end{equation} 
 This 2D Lorentzian manifold having co-ordinates $(a, \phi)$ is termed as minisuperspace. On the otherhand, the dynamical part is defined by the potential 
 \begin{equation}
V_{eff} (a, \phi)=2a^3 \Bigg[V(\phi)+\rho_0 a^{-3\gamma}\Bigg].\label{3.13}
 \end{equation} 
In quantum cosmology, the wave function of the Universe is determined by solving the WD equation which is a 2nd order hyperbolic partial differential equation. In fact WD equation is the Klein Gordon (KG) equation defined by the conformal Laplacian operator over the minisuperspace as
\begin{equation}
\bigtriangleup \psi +\frac{(n-2)}{4(n-1)}R(x^k) \psi +V_{eff}(x^k) \psi=0,\label{3.14}
 \end{equation}
 where $\bigtriangleup=\frac{1}{\sqrt{|g|}}\frac{\partial}{\partial x^i}\big(\sqrt{|g|}\frac{\partial}{\partial x^j}\big)$ is the conformal Laplacian operator, $g_{ij}$ is the metric and $n$ is the dimension of the minisuperspace. So in the present minisuperspace (having dimension $n=2$) the WD equation becomes
 \begin{equation}
 \bigtriangleup \psi +2a^3 \Bigg[V(\phi)+\rho_0 a^{-3\gamma}\Bigg]\psi=0,\label{3.15}
 \end{equation}
 with the Laplacian operator defined by:
 \begin{equation}
 \bigtriangleup\equiv -\frac{1}{6a} \Bigg(\frac{\partial^2}{\partial a^2}+\frac{\partial}{\partial a}\Bigg)+\frac{1}{a^3 \lambda(\phi)}\frac{\partial^2}{\partial \phi^2}.\label{3.16}
 \end{equation}
Paliathanasis and Tsamparlis \cite{r21.1} showed that the Lie point symmetries of the KG equation (\ref{3.14}) are connected to the conformal algebra of the minisuperspace metric $g_{ij}$. According to them the general form of the Lie point symmetry vector can be expressed as  
\begin{equation}
\overrightarrow{X}=\xi^i (x^k)\partial_i
\Bigg[\frac{(2-n)}{2}\lambda \psi+a_0 \psi\Bigg] \partial \psi,\label{3.17}
\end{equation}
 where $\xi^i$ is a conformal killing vector of the minisuperspace having the conformal factor $\lambda(x^k)$. Also the Lie point symmetry condition restricts the potential to
 \begin{equation}
 L_{\overrightarrow{\xi}}V_{eff}+2 \lambda V_{eff}=0.\label{3.18}
 \end{equation}
 Here $L_{\overrightarrow{\xi}}$ is the above symmetry vector $\overrightarrow{X}$ (in equation (\ref{3.17})) which can be simplified a bit by the coordinate transformation $x^i \rightarrow u^i$ so that $\xi^i \partial_i \rightarrow \frac{\partial}{\partial u_j}$ such that $\overrightarrow{X}$ becomes
 \begin{equation}
\overrightarrow{X}=\partial_j+\Bigg[\frac{(2-n)}{2}\lambda \psi+a_0 \psi\Bigg]\partial \psi,\label{3.19}
 \end{equation}
 One can now reduce the WD equation using this symmetry vector in the following two equivalent ways:
 
 {\bf I.} In this approach considering the symmetry vector $\overrightarrow{X}$ as a Lagrange system, the zero-order invariants give
 \begin{equation}
 \frac{dy^I}{0}= \frac{dy^J}{1}= \frac{d\psi}{\Bigg(\frac{2-n}{2} \lambda+a_0\Bigg)\psi},\label{3.20}
 \end{equation}
 (with $I \neq J$) which has the solution
 \begin{eqnarray}
 y^I&=&C^I,~\mbox{a constant},\nonumber\\
 \mbox{and}~\psi( y^I,  y^J)&=&\psi_0 ( y^I) \exp\Bigg[\int\Bigg\{\Bigg(\frac{2-n}{2}\Bigg)\lambda+a_0\Bigg\}dy^J\bigg].\label{3.21}
 \end{eqnarray}
 
  {\bf II.} Alternatively, one can extend the above Lie point symmetry to Lie-B$\ddot{\mbox{a}}$cklund symmetry so that the symmetry (\ref{3.19}) becomes the contact symmetry
  \begin{equation}
  \tilde{X}=\Bigg[\psi_{,J}-\Bigg(\frac{2-n}{2}\lambda+a_0\Bigg)\psi\Bigg]\partial_{\psi}.\label{3.22}
  \end{equation}
  
  Hence the differential equation for $\psi$ now becomes
   \begin{eqnarray}
   \psi_{,J}&-&\Bigg(\frac{2-n}{2}\lambda+a_0\Bigg)\psi=a_1 \psi,\nonumber
   \end{eqnarray}
   that is
   \begin{eqnarray}
   \psi_{,J}&=&\Bigg(\frac{2-n}{2}\lambda+b_0\Bigg)\psi,~b_0=a_0+a_1\label{3.23}
   \end{eqnarray}
 which has the solution
 \begin{equation}
 \psi( y^I,  y^J)=\psi_1 ( y^I) \exp\Bigg[\int \Bigg\{\Bigg(\frac{2-n}{2}\Bigg)\lambda+b_0\Bigg\}dy^J\bigg].\nonumber
 \end{equation}
 
 On the otherhand, in the context of WKB approximation one can write the wave function as $\psi(x^k) \sim e^{is}(x^k)$ and consequently the WD equation (\ref{3.15}) reduces to the (null) Hamilton-Jacobi equation:
 \begin{equation}
 -\frac{1}{12a} \Bigg(\frac{\partial s}{\partial a}\Bigg)^2+\frac{1}{2a^3 \lambda(\phi)} \Bigg(\frac{\partial s}{\partial \phi}\Bigg)^2+a^3 V(\phi)+\rho_0 a^{-3(\gamma-1)}=0.\label{3.24}
 \end{equation}
 
 Note that this first order non-linear partial differential equation is nothing but the Hamilton-Jacobi (H--J) equation of a Hamiltonian system moving in the same geometry under the conformal Laplace operator of the WD equation having the same potential. Thus to obtain the invariant solution of the WD equation and the Hamiltonian system to be Liouville integrable there should be at-least $(n-1)$ independent Lie point symmetries (which form an Abelian Lie algebra) of a $n-$dimensional WD equation. So the solution of the WD equation can be expressed in terms of zero order invariants of these Lie point symmetries as
 \begin{equation}
 \psi(\tilde{x}^n, \tilde{x}^J)=\phi (\tilde{x}^n) \exp \Bigg[ \sum^{n-1}_{J=1} \int \Bigg\{\Bigg(\frac{2-n}{2}\Bigg)\lambda-Q_J\Bigg\}d\tilde{x}^J\Bigg],\nonumber 
 \end{equation} 
 where $Q_{_J}$'s are the constants of motion along the symmetry directions and $\phi(\tilde{x}^n)$ satisfies a linear second order ordinary differential equation. Thus for reduction or solution of the WD equation, Lie point symmetries of the WD equation can be considered and these symmetries can be determined using the conformal killing vector of the minisuperspace.\\
 
 In section III it has been shown that the conformal killing algebra formed by the conformal killing vectors are different for two conformally related metrics. As a result the Lie symmetry vector which reduces/solves the WD equation changes due to conformal transformation and hence the above solutions will be different.
 
 \section{A General study of Noether symmetry}  
 
 From the point of view of general methods to have conserved quantities, the Noether symmetry approach will be very much relevant. A vector field $\overrightarrow{X}$ defined over the tangent space of configurations $TQ\equiv \{q, \dot{q}\}$ can be written as
 \begin{equation}
 \overrightarrow{X}=\alpha(q) \frac{\partial}{\partial q}+\dot{\alpha}(q) \frac{\partial}{\partial \dot{q}}.\label{q12}
 \end{equation}
 Then according to Noether theorem \cite{n6},\cite{n7}
 \begin{eqnarray}
 L_{ \overrightarrow{X}} \mathcal{L}=0,\nonumber
 \end{eqnarray}
 that is
 \begin{eqnarray}
  \overrightarrow{X} \mathcal{L}=\alpha \frac{\partial \mathcal{L}}{\partial q}+\dot{\alpha}\frac{\partial \mathcal{L}}{\partial \dot{q}},\label{q13}
 \end{eqnarray}
 where the Lagrangian $\mathcal{L}$ is defined over the tangent space of configurations (i.e., $TQ$). The above condition corresponds to a constant of motion for the Lagrangian i.e., the phase flux is conserved along the vector field $\overrightarrow{X}$. On the otherhand, in the case of the Hamiltonian formulation, the above (Noether) symmetry condition becomes
  \begin{equation}
  L_{ \overrightarrow{v}} \mathcal{H}=0,\label{q14}
 \end{equation}
 with $ \overrightarrow{v}=\dot{q}\frac{\partial}{\partial q}+\ddot{q} \frac{\partial}{\partial \dot{q}}$.
 
 We shall now apply this symmetry condition to the minisuperspace models of quantum cosmology to obtain appropriate interpretation of the wave function of the Universe. Due to Noether symmetry one gets the conserved canonically conjugate momenta as
 \begin{equation}
 \Pi_i \equiv \frac{\partial\mathcal{L}}{\partial q^i}=i_{\overrightarrow{x_i}} \theta_{\mathcal{L}}=\Sigma_i,~i=1, 2,.....m\label{q15}
 \end{equation}
 where `$m$' is the number of symmetries. So on quantization for each $i$
 \begin{equation}
 -i \partial_{q^i}\big|\psi\big>=\Sigma_i \big|\psi\big>,\label{q16}
 \end{equation}
 which implies that a translation along the $q^i$- axis is singled out by the corresponding symmetry. Assuming the conserved quantities $\sum_i$ to be real one gets oscillatory solution components for the wave function along the symmetries as 
 \begin{equation}
 \big|\psi\big>=\sum^m_{k=1} e^{i \sum_k q^k}\big|\phi(q^l)\big>,~l<n, \label{q17} 
 \end{equation}
 where the index `$l$' stands for directions along which there is no symmetry and `$n$' is the dimension of the minisuperspace. Thus for the existence of Noether symmetry, the wave function has an oscillatory part and the conjugate momenta along the symmetry directions should be conserved.
 
 Further, due to symmetries there are first integrals of motion and it is possible to choose classical trajectories. In particular, if the minisuperspace is of dimension one or two then due to Noether symmetry one can have complete solution of the problem and as a result there is full semi-classical limit of quantum cosmology. Then the Noether symmetries and the corresponding reduction procedure of dynamics selects a subset having oscillatory behaviour of the solution of the WD equation. According to Hartle \cite{r36}, the wave function of the Universe with the above symmetry condition corresponds to conserved momenta and the trajectories can be interpreted as classical cosmological solutions. Conversely, if a subset of the solution of the WD equation has an oscillatory behaviour then there exists conserved momenta and Noether symmetries along those directions.
 
 \section{Noether symmetry in Non-minimally coupled scalar field cosmology}
 The Lagrangian is defined on the 2D configuration space $\{a, \phi\}$ (which is termed as minisuperspace) and hence the Noether symmetry vector field
 \begin{equation}
 \overrightarrow{X}=\alpha\frac{\partial}{\partial a}+\beta\frac{\partial}{\partial \phi}+\dot{\alpha}\frac{\partial}{\partial \dot{a}}+\dot{\beta}\frac{\partial}{\partial \dot{\phi}},\label{5.1}
 \end{equation}
 acts on this configuration space. Here $\alpha=\alpha(a,\phi),~~\beta=\beta(a,\phi),~~\dot{\alpha}=\frac{\partial \alpha}{\partial a}\dot{a}+\frac{\partial \alpha}{\partial \phi}\dot{\phi},~\mbox{and}~\dot{\beta}=\frac{\partial \beta}{\partial a}\dot{a}+\frac{\partial \beta}{\partial \phi}\dot{\phi}.$
 
 Now the Noether symmetry condition (i.e., $L_{\overrightarrow{X}} \mathcal{L}=0$) results in a system of partial differential equations
 \begin{equation}
 \alpha+2a\frac{\partial \alpha}{\partial a}=0,\label{5.2}
 \end{equation}
 \begin{equation}
 6\frac{\partial \alpha}{\partial \phi}-a^2\lambda(\phi)\frac{\partial \beta}{\partial a}=0,\label{5.3}
 \end{equation}
 \begin{equation}
 3\alpha\lambda+\beta a\lambda^\prime+2a\lambda\frac{\partial \beta}{\partial \phi}=0,\label{5.4}
 \end{equation}
 and
 \begin{eqnarray}
 3\alpha\rho_0+3\alpha V(\phi)+\beta aV^\prime(\phi)&=&0,~~\mbox{for}~~\gamma=0,\nonumber\\
 3\alpha V(\phi)+\beta aV^\prime(\phi)&=&0,~~\mbox{for}~~\gamma=1.\label{5.5}
 \end{eqnarray}
 Using the method of separation of variables one writes $\alpha=\alpha_1(a)\alpha_2(\phi)~,~\beta=\beta_1(a)\beta_2(\phi)$ and choosing $\lambda(\phi)=\frac{\lambda_0}{\phi^2}$  (from the Lie symmetry of the evolution equations, for details see reference \cite{r24}) the solutions are the following:\\

 {\bf A: $\lambda_0 > 0$}
 \begin{equation}
 \alpha=\frac{A_0}{\sqrt{a}}\cosh(p~\ln~\phi+b_1),\label{5.6}
 \end{equation}
 \begin{equation}
 \beta=-\frac{4A_0}{\kappa^2 \lambda_0}pa^{-\frac{3}{2}}\phi \sin h(p~\ln~\phi+b_1),\label{5.7}
 \end{equation}
 \begin{equation}
 V=\left\{
 \begin{array}{ll}
 V_0\sinh^2(p~\ln~\phi+b_1), & \mbox{ for } \gamma=1\\
 V_0\sinh^2(p~\ln~\phi+b_1)-\rho_0, & \mbox{ for } \gamma=0
 \end{array}
 \right.\label{5.8}
 \end{equation}
 
 {\bf B: $\lambda_0 < 0$}
 \begin{equation}
 \alpha=\frac{A_0}{\sqrt{a}}cos(p~\ln~\phi+b_1),\label{5.9}
 \end{equation}
 \begin{equation}
 \beta=-\frac{4A_0}{\kappa^2 \left|\lambda_0\right|}pa^{-\frac{3}{2}}\phi \sin(p~\ln~\phi+b_1),\label{5.10}
 \end{equation}
 \begin{equation}
 V=\left\{
 \begin{array}{ll}
 V_0\sin^2(p~\ln~\phi+b_1), & \mbox{ for } \gamma=1\\
 V_0\sin^2(p~\ln~\phi+b_1)-\rho_0, & \mbox{ for } \gamma=0
 \end{array}
 \right.,\label{5.11}
 \end{equation}
 with $p^2=\frac{3}{8}\kappa^2\left|\lambda_0\right|$,~$A_0$, an arbitrary integration constant.
 
 On the otherhand, assuming the potential of the scalar field to be exponential i.e., $V=V_0e^{\mu \phi},~ (V_0, \mu$ are constants) the above set of partial differential equations has the solutions:

 ${\bf C.~\gamma=0}$
 \begin{eqnarray}
 \alpha&=&\alpha_0a^{\frac{-1}{2}}\left[e^{\mu \phi}\left(1+\frac{\rho_0}{V_0}e^{-\mu \phi}\right)-\frac{\rho_0}{V_0}ln\left(e^{\mu \phi}+\frac{\rho_0}{V_0}\right)\right]^\frac{1}{2},
 \nonumber
 \\
 \beta&=&-\beta_0a^\frac{-3}{2}\left(1+\frac{\rho_0}{V_0}e^{-\mu \phi}\right)\left[e^{\mu \phi}\left(1+\frac{\rho_0}{V_0}e^{-\mu \phi}\right)-\frac{\rho_0}{V_0}ln\left(e^{\mu \phi}+\frac{\rho_0}{V_0}\right)\right]^\frac{1}{2},
 \nonumber
 \\
 \lambda&=&\lambda_0\frac{e^{\mu \phi}}{\left(1+\frac{\rho_0}{V_0}e^{-\mu \phi}\right)^2\left[e^{\mu \phi}\left(1+\frac{\rho_0}{V_0}e^{-\mu \phi}\right)-\frac{\rho_0}{V_0}ln\left(e^{\mu \phi}+\frac{\rho_0}{V_0}\right)\right]^\frac{1}{2}}.\label{5.12}
 \end{eqnarray}
 
 ${\bf D.~\gamma=1}$
 \begin{eqnarray}
 \alpha&=&\alpha_0a^\frac{-1}{2}\left[b_0+c_0e^{\mu \phi}\right]^\frac{1}{2},
 \nonumber
 \\
 \beta&=&-\beta_0a^\frac{-3}{2}\left[b_0+c_0e^{\mu \phi}\right]^\frac{1}{2},
 \nonumber
 \\
 \lambda&=&\lambda_0\frac{e^{\mu \phi}}{\left[b_0+c_0e^{\mu \phi}\right]},\label{5.13}
 \end{eqnarray}
 with  $\alpha_0~,\beta_0~,\lambda_0~,b_0$ and $c_0$ being arbitrary constants.
 
 To simplify the Lagrangian using cyclic variables the transformation of the minisuperspace variables take the following forms:
 
 {\bf For: $\lambda_0 > 0$}
 \begin{eqnarray}
 a&=&\left(\frac{3A_0}{2}\right)^{\frac{2}{3}}\left(u^2-v^2\right)^{\frac{1}{3}},\nonumber\\
 \phi&=&\exp\left[\frac{1}{p}\left(\tanh^{-1}(\frac{v}{u})-b_1\right)\right].\label{5.14}
 \end{eqnarray}
 
 {\bf For: $\lambda_0 < 0$}
 \begin{eqnarray}
 a&=&\left(\frac{3A_0}{2}\right)^{\frac{2}{3}}\left(u^2+v^2\right)^{\frac{1}{3}},\nonumber\\
 \phi&=&\exp\left[\frac{1}{p}\left(\tan^{-1}(\frac{v}{u})-b_1\right)\right].\label{5.15}
 \end{eqnarray}
 
 As a result the transformed Lagrangian takes the simple form 
 
\begin{equation}
L=\left\{
\begin{array}{llll}
3A_0^2(\dot{u}^2-\dot{v}^2)+\frac{9}{4}\kappa^2A_0^2V_0v^2+\kappa^2\rho_0 & \mbox{ for } \lambda_0{>}0,\gamma=1\\
3A_0^2(\dot{u}^2-\dot{v}^2)+\frac{9}{4}\kappa^2A_0^2V_0v^2 & \mbox{ for } \lambda_0{>}0,\gamma=0\\
3A_0^2(\dot{u}^2+\dot{v}^2)+\frac{9}{4}\kappa^2A_0^2V_0v^2+\kappa^2\rho_0 & \mbox{ for } \lambda_0{<}0,\gamma=1\\
3A_0^2(\dot{u}^2+\dot{v}^2)+\frac{9}{4}\kappa^2A_0^2V_0v^2 & \mbox{ for } \lambda_0{<}0,\gamma=0
\end{array}
\right.,\label{5.16}
\end{equation}
where the variable `$u$' is cyclic in nature.

The conjugate momenta corresponding to the new variables $(u, v)$ are
\begin{equation}
\Pi_u=\frac{\partial L}{\partial \dot{u}}=6A_0^2 \dot{u},~\mbox{for all} ~ \gamma~\mbox{and}~ \lambda_0\nonumber 
\end{equation}

\begin{equation}
\Pi_v=\frac{\partial L}{\partial \dot{v}}=\left\{
\begin{array}{ll}
 -6A_0^2 \dot{v} & \mbox{ for } \lambda_0{>}0, ~\mbox{for all}~\gamma\\
+6A_0^2 \dot{v} & \mbox{ for } \lambda_0{<}0, ~\mbox{for all}~\gamma
\end{array}
\right..\nonumber
\end{equation}

Also the corresponding Hamiltonian in the new variables takes the form
\begin{equation}
\mathcal{H}=\left\{
\begin{array}{llll}
\frac{1}{12A_0^2}(\Pi_{u}^2-\Pi_{v}^2)-\frac{9}{4}\kappa^2A_0^2V_0v^2-\kappa^2\rho_0, & \mbox{ for } (\lambda_0{>}0, \gamma=1)\\
\frac{1}{12A_0^2}(\Pi_{u}^2-\Pi_{v}^2)-\frac{9}{4}\kappa^2A_0^2V_0v^2, & \mbox{ for } (\lambda_0{>}0, \gamma=0)\\
\frac{1}{12A_0^2}(\Pi_{u}^2+\Pi_{v}^2)-\frac{9}{4}\kappa^2A_0^2V_0v^2-\kappa^2\rho_0, & \mbox{ for } (\lambda_0{<}0,~\gamma=1)\\
\frac{1}{12A_0^2}(\Pi_{u}^2+\Pi_{v}^2)-\frac{9}{4}\kappa^2A_0^2V_0v^2, & \mbox{ for } (\lambda_0{<}0, \gamma=0)
\end{array}
\right..\label{5.17}
\end{equation}
The Noether symmetry in the new variables is given by
\begin{eqnarray}
\Pi_u=6A_0^2 \dot{u}=\Sigma_0.\label{5.17.1}
\end{eqnarray}
For canonical quantization we write the operator version: $\Pi_u \rightarrow -i \partial_u,~ \Pi_v \rightarrow -i \partial_v$ and the Wheeler-DeWitt equation now becomes
\begin{eqnarray}
\Bigg[\frac{\partial^2}{\partial u^2}-\frac{\partial^2}{\partial v^2}+3A_0^2 \kappa^2\Bigg(9A_0^2V_0v^2+\rho_0\Bigg)\Bigg]\Psi=0,~\mbox{for}~(\lambda_0{>}0, \gamma=1)\nonumber\\
\Bigg[\frac{\partial^2}{\partial u^2}-\frac{\partial^2}{\partial v^2}+27A_0^4 \kappa^2 V_0v^2\Bigg]\Psi=0,~\mbox{for}~(\lambda_0{>}0, \gamma=0)\nonumber\\
\Bigg[\frac{\partial^2}{\partial u^2}+\frac{\partial^2}{\partial v^2}+37A_0^2 \kappa^2\Bigg(9A_0^2V_0v^2+\rho_0\Bigg)\Bigg]\Psi=0,~\mbox{for}~(\lambda_0{<}0, \gamma=1)\nonumber\\
\Bigg[\frac{\partial^2}{\partial u^2}+\frac{\partial^2}{\partial v^2}+27A_0^4 \kappa^2 V_0v^2\Bigg]\Psi=0,~\mbox{for}~(\lambda_0{<}0, \gamma=0)\label{5.17.2}
\end{eqnarray}

Also the quantum version of the constraint equation takes the form,
\begin{equation}
-i \partial_u\big|\Psi\big>=\Sigma_0\big|\Psi\big>\label{5.17.3}
\end{equation}                                                                                                                                                                                                   which has the simple solution: $\big|\Psi\big>=\big|\Phi(v)\big> \exp (i \Sigma_0 u)$.

So from the WD equation $\big|\Phi(v)\big>$ satisfies
\begin{eqnarray}
	\Bigg[\frac{d^2}{d v^2}+\Bigg(\xi^2-\mu^2 v^2\Bigg)\Bigg]\big|\Phi(v)\big>=0,~\mbox{for}~\lambda_0{>}0, \gamma=1\nonumber\\
	\Bigg[\frac{d^2}{d v^2}+\Bigg(\xi_0^2-\mu^2 v^2\Bigg)\Bigg]\big|\Phi(v)\big>=0,~\mbox{for}~\lambda_0{>}0, \gamma=0\nonumber\\
	\Bigg[\frac{d^2}{d v^2}+\Bigg(\mu^2 v^2-\xi^2\Bigg)\Bigg]\big|\Phi(v)\big>=0,~\mbox{for}~\lambda_0{<}0, \gamma=1\nonumber\\
\Bigg[\frac{d^2}{d v^2}+\Bigg(\mu^2 v^2-\xi_0^2\Bigg)\Bigg]\big|\Phi(v)\big>=0,~\mbox{for}~\lambda_0{<}0, \gamma=0\label{5.17.4}
\end{eqnarray}
with $\mu^2=27 \kappa^2 A_0^4 V_0,~\xi^2=\Sigma_0^2-3A_0^2\kappa^2 \rho_0,~\xi_0^2=\Sigma_0^2.$\\
 The first two equations in equation (\ref{5.17.4}) are nothing but analogous to the time independent Schr$\ddot{\mbox{o}}$dinger equation for quantum harmonic oscillator while the last two differential equations in equation (\ref{5.17.4}) have solutions as parabolic cylinder functions. Thus the solutions of the first two equations in (\ref{5.17.4}) are time independent wave function of the quantum harmonic oscillator in terms of Hermite polynomials which may be written as 
	\begin{equation}
	\big|\Phi(v)\big>=\bigg(\frac{\mu}{\Pi}\bigg)^{\frac{1}{4}}\frac{1}{\sqrt{2^n n!}}e^{-\frac{\mu v^2}{2}}H_n(\sqrt{\mu}v),~n=0, 1, 2,....
	\end{equation}\label{67}
	where $H_n(x)$ is a Hermite polynomial of degree `$n$'. Note that the parameter $\mu$ is related to the angular frequency of the oscillator as $\mu=\frac{mw}{\hbar}$ and the energy eigenvalues are related to the parameter $\xi$ (or $\xi_0$) by the relation $\xi^2=\frac{2mE}{\hbar^2}$. Similarly, the solutions of the other two differential equations can be written in terms of Whitaker function with complex argument, for details see reference \cite{r37}.
The solutions of the above WD equation becomes:\\
{\bf For : $\lambda_0 >0, \gamma=1$}

\begin{equation}
\phi(v)=\frac{C_1\mbox{Whitaker M}\bigg(\frac{1}{4} \frac{\xi^2}{\mu},\frac{1}{4}, v^2 \mu \bigg)}{\sqrt{v}}+\frac{C_2\mbox{Whitaker W}\bigg(\frac{1}{4} \frac{\xi^2}{\mu},\frac{1}{4}, v^2 \mu \bigg)}{\sqrt{v}}\label{68}
\end{equation}

{\bf For : $\lambda_0 <0, \gamma=1$}

\begin{equation}
\phi(v)=\frac{C_1\mbox{Whitaker M}\bigg(\frac{\frac{1}{4}I \xi^2}{\mu},\frac{1}{4}, I v^2 \mu \bigg)}{\sqrt{v}}+\frac{C_2\mbox{Whitaker W}\bigg(\frac{\frac{1}{4}I \xi^2}{\mu},\frac{1}{4}, I v^2 \mu \bigg)}{\sqrt{v}}
\end{equation}\label{69}
\begin{figure}
	\centering
	\includegraphics[width=0.45\textwidth]{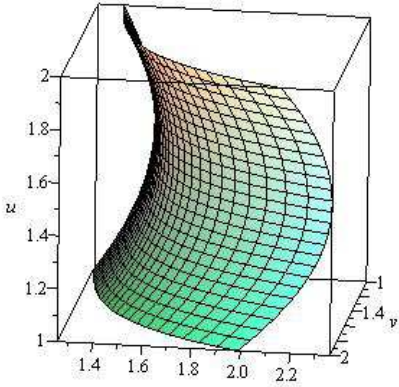}\\
	\caption{Behaviour of the wave function}
	\label{fig1}
\end{figure}

\section{Brief Discussion and concluding remarks}
This work presents an extensive study of quantum cosmology for non-minimally coupled scalar field in homogeneous and isotropic FLRW geometry from the perspective of symmetry analysis. For the present physical model the minisuperspace is a two dimensional Lorentzian manifold and WD equation is constructed on this minisuperspace. The Lie point symmetries of this WD equation are shown to be related to the conformal algebra of the metric of the minisuperspace and the symmetry vector is found to be related to the conformal killing vector of the minisuperspace. The study of Noether symmetries to the minisuperspace gives oscillatory solution to the WD equation and is nothing but the semi-classical limit of quantum cosmology. Also the Noether conserved current is associated with the conserved momenta of the system. Finally, the non-oscillatory part of the WD equation is found to be time-independent Schr$\ddot{\mbox{o}}$dinger equation for harmonic oscillator and the corresponding analytic solutions are presented. Also the graphical representation of the wave function of the Universe (from equation (\ref{5.17.3}) and (\ref{68})) is shown in Fig.\ref{fig1}.\\

Usually, for non-minimally coupled scalar field cosmology, one redefines the scalar field so that the physical problem reduces to minimally coupled scalar field cosmology. Essentially, this is nothing but a conformal transformation of the minisuperspace metric. As the symmetry vector corresponding to Lie symmetry is related to the conformal killing vector of the minisuperspace metric while that for Noether symmetry it is related to the homothetic vector field so that after conformal transformation, although the conformal algebra remains the same, the two subalgebras namely killing algebra and homothetic algebra will be different. Hence the quantum cosmology so constructed will not be identical due to conformal transformation. Finally, one may conclude that symmetry analysis of the minisuperspace has a great role in quantum cosmology, particularly in solving the WD equation. 

\section*{Acknowledgments}
Author SD acknowledges Science and Engineering Research Board (SERB), Govt. of India, for
awarding a National Post-Doctoral Fellowship (File No: PDF/2016/001435) and the Department of
Mathematics, Jadavpur University where a part of the work was completed. Author ML thanks the
Department of Science and Technology, Government of India for the award of a DST-SERB Distinguished Fellowship. S.C. thanks Science and Engineering Research Board (SERB) for awarding
MATRICS Research Grant support (File No. MTR/2017/000407) and Inter
University Center for Astronomy and Astrophysics (IUCAA), Pune, India for their
warm hospitality as a part of the work was done during a visit.

\end{document}